\documentclass[namedreferences]{SolarPhysics}
\usepackage[optionalrh]{spr-sola-addons}
\usepackage{solaheader} 
\usepackage{graphicx}        
\usepackage{color}           
\usepackage{url}             

\begin{document}

\begin{article}

\begin{opening}

\title{Quasi-Periodic Releases of Streamer Blobs and Velocity Variability of the Slow Solar Wind near the Sun}

\author{H.Q. \surname{Song}$^{1,2}$\sep
        Y. \surname{Chen}$^{1}$\sep
        K. \surname{Liu}$^{3}$\sep \\
        S.W. \surname{Feng}$^{1}$\sep
        L.D. \surname{Xia}$^{1}$
       }
\runningauthor{H.Q. Song, \textit{et al.}}
\runningtitle{Quasi-Periodic Releases of Streamer Blobs}

   \institute{$^{1}$ School of Space Science and Physics, Shandong University at Weihai, Weihai Shandong 264209,
                     China\\
                     email: \url{yaochen@sdu.edu.cn}\\
             $^{2}$ State Key Laboratory of Space Weather, Chinese Academy of Sciences, Beijing 100190,
                     China\\
             $^{3}$ School of Earth and Space Sciences, University of Science and Technology of China, Hefei Anhui 230026, China
             }

\begin{abstract}
We search for persistent and quasi-periodic release events of
streamer blobs during 2007 with the Large Angle Spectrometric
Coronagraph on the \textit{Solar and Heliospheric Observatory} and
assess the velocity of the slow solar wind along the plasma sheet
above the corresponding streamer by measuring the dynamic
parameters of blobs. We find 10 quasi-periodic release events of
streamer blobs lasting for three to four days. In each day of
these events, we observe three-five blobs. The results are in line
with previous studies using data observed near the last solar
minimum. Using the measured blob velocity as a proxy for that of
the mean flow, we suggest that the velocity of the background slow
solar wind near the Sun can vary significantly within a few hours.
This provides an observational manifestation of the large velocity
variability of the slow solar wind near the Sun.
\end{abstract}
\end{opening}

\section{Introduction}

Sheeley \textit{et al}. (1997) were the first to report the
observation of plasma blobs, released from the tips of streamers
as revealed in the data obtained by the Large Angle Spectrometric
Coronagraph (LASCO) on the \textit{Solar and Heliospheric
Observatory} (SOHO) spacecraft (Brueckner \textit{et al}., 1995)
around the last solar minimum. According to the data analysis by
Sheeley \textit{et al}. (1997) and a series of following studies
by Wang \textit{et al}. (1998,2000) blobs emerge at about 2-4
Solar Radii ($R_\odot$) from the Sun center as radially elongated
structures with initial sizes being about 1 $R_\odot$ in the
radial direction and 0.1 $R_\odot$ in the transverse direction.
They move outward radially, maintaining an almost constant angular
span and their lengths increased from $\approx$1 $R_\odot$ to
$\approx$3 $R_\odot$ within the LASCO field of view (FOV). Their
velocities also increase gradually with increasing length. Besides
understanding the plasma process accounting for the formation of
blobs themselves, there are at least two other issues directly
related to blob studies: The first is that blob studies provide a
practical technique of assessing the velocity of the embedded
solar wind, since blobs are believed to get closely coupled to and
flow outward together with the background solar wind shortly after
their emission. This component of the solar wind (\textit{i.e.},
the wind originated from the plasma sheet above a streamer) is
usually taken to be a part of the slow solar wind (\textit{e.g.},
Woo and Martin, 1997; Sheeley \textit{et al}., 1997; Habbal
\textit{et al}., 1997; Wang \textit{et al}., 2000). That is to say
that the measurements of the blob motion can be used to represent
the velocity of the embedded slow wind plasmas above a certain
distance. Another issue of blob studies concerns the possible
diagnostics of plasma properties enclosed in the closed
magnetic-field regions of streamers, especially, in the streamer
cusp region, through \textit{in situ} detection of blob structures
in interplanetary space. This is deduced from the assumption that
blobs originate from inside the closed arcades right below the
streamer cusp, which is used or supported by several physical and
numerical models of blob formation (\textit{e.g.}, Wu \textit{et
al}., 2000; Wang \textit{et al}., 2000; Chen \textit{et al}.,
2009). Note that there are also models that suggest that blobs are
the aftermath of magnetic reconnections along the current sheet
embedding in the solar wind with open magnetic geometry
(\textit{e.g.}, Einaudi \textit{et al}., 1999; Lapenta and Knoll,
2005). The possibility of collecting samples of plasmas originated
from inside the closed magnetic-field regions with \textit{in
situ} measurements is important for the evaluation of elemental
compositions in the blob source region, and for understanding the
formation and stability of coronal streamers, as well as the
delicate coupling process between plasmas and magnetic field near
the cusps.

Wang \textit{et al}. (1998) reported a very interesting event with
steady quasi-periodic releases of blobs above a streamer during
the eight days from 19 to 26 April, 1997. The daily rate of blobs
in this event is observed to be three-five with a release period
ranging from five to eight hours. To interpret the formation and
quasi-periodic releases of blobs, Chen \textit{et al}. (2009)
designed a numerical model accounting for the magnetohydrodynamic
coupling process between the closed streamer magnetic arcades and
the solar wind expansion. They found that the streamer-cusp
geometry is subject to an intrinsic instability originating from
the peculiar magnetic topological feature at the cusp region
despite the long-term stability of the overall morphology.
According to Chen \textit{et al}. (2009), the process of the
instability consists of two successive processes. One is the
plasma magnetic-field expansion through the localized cusp region
where the field is too weak to maintain plasma confinement; the
continuing expansion brings strong velocity shear into the slow
wind regime, providing the free energy necessary for the onset of
a streaming sausage mode instability (Lee, Wang, and Wei, 1988;
Wang, Li, and Wei, 1988). The other is then the onset and
nonlinear development of the streaming instability, which causes
pinches of magnetic-field lines and drives reconnections at the
pinching points to form separated magnetic blobs. After the birth
of a blob, the streamer system returns to the configuration with a
lower cusp point, subject to another cycle of the instability. As
mentioned, the whole process originates from the topological
feature at the cusp region, which is intrinsically associated with
a typical coronal streamer; therefore Chen \textit{et al}. (2009)
made use of the word ``intrinsic'' to describe the streamer
instability. We point out in passing that other numerical models
demonstrating various aspects of streamer instabilities exist in
the literature (\textit{e.g.}, Suess, Wang, and Wu, 1996; Wu
\textit{et al}., 2000; Endeve, Leer, and Holzer, 2003; Endeve,
Holzer, and Leer, 2004). According to the numerical results given
by Chen \textit{et al}. (2009), the period of blob formation is
about four-five hours. Thus, hypothetically, one can observe
four-six blobs per day on average, in agreement with what is
observed by Wang \textit{et al}. (1998). We find this agreement
with the observations to be very encouraging considering the
simplicity of Chen \textit{et al}.'s numerical model (2009).
However, in the series of blob studies by Wang \textit{et al}.
(1998,2000), only a few events with continuous and quasi-periodic
releases of blobs are reported (in April 1997 and December 1998).
If the scenario proposed by Chen \textit{et al}. (2009) is
basically correct-that the blobs are the aftermath of an intrinsic
instability of coronal streamers with release periods being
several hours-there should exist more events with steady blob
releases. It is the primary purpose of this paper to search for
events similar to those reported by Wang \textit{et al}. (1998)
with the LASCO data. As a starting point, we only deal with the
data accumulated in the whole year of 2007 in this paper.

As mentioned previously, the velocity measurement of blobs can be
used as a proxy for that of the embedded slow solar wind along the
plasma sheet. This argument is further supported by a recent
numerical calculation presented in Chen \textit{et al}. (2009).
They show that, as a result of the dynamical coupling to the mean
flow, the blobs are basically accelerated to the same velocity
after they further propagate a distance of 2-3 $R_\odot$ from the
disconnection point. Therefore, in general, beyond a certain
heliocentric distance of, say, 5 to 7 $R_\odot$, the background
solar wind velocity can be well represented by that of the blobs.
However, most blobs are too weak to be observable beyond 20
$R_\odot$ by the LASCO C3 coronagraph. Therefore, the major region
where this method is usable is limited to 4-20 $R_\odot$. At
present, there are only a few other indirect techniques, such as
the Doppler dimming technique (\textit{e.g.}, Li \textit{et al}.,
1998; Cranmer \textit{et al}., 1999; Strachan \textit{et al}.,
2002) and the IPS (Interplanetary Scintillation) technique
(\textit{e.g.}, Grail \textit{et al}., 1996; Breen \textit{et
al}., 1999), that can be used to determine the wind velocity
within the first 20 $R_\odot$ of the corona. For instance, one can
use the measured ratio of the O VI doublet to evaluate the
outflowing velocities of O$^{5+}$ ions. The velocities obtained by
both the Doppler dimming technique and the IPS techniques are
usually model dependent with large errors. As mentioned, the
presence of blobs provides another velocity diagnostic technique
of the solar wind in the corona. This can be referred to as the
blob technique. Among the various methods of velocity measurement
in the corona, the blob technique may serve as the most accurate,
at least in cases where blobs are clearly measurable. One serious
limitation of this method is that only the projected velocity of
the solar wind along the plasma sheet can be revealed. Also, it
should be noted that the blob technique is based on the general
assumption that blobs can be taken as effective velocity tracers
of the mean flow. Nevertheless, the second purpose of this paper
is to examine the velocity of the solar wind along the plasma
sheet, which is usually regarded as a part of the source region of
the slow solar wind, as mentioned previously. The details of our
observations and results are described in the following section.
The summary and discussion are provided in the last section of
this paper.

\section{Observations and Results}

As already mentioned, one of the main purposes of this paper is to
search for steady release events of blobs. To investigate the
quasi-periodic character of blob emissions, we need to observe
enough blobs emitted above a streamer. Therefore, only those
events with emission lasting for at least three days are reported
in this study. By examining all of the white-light data taken by
the LASCO coronagraph in 2007, we have identified 10 events with
steady emission of blobs lasting for three to four days. Some
information about these 10 events is listed in Table 1, where the
number in the first column indicates the time sequence of the blob
emission. In the remaining columns, we list the start and end
dates of the events, the position angle (PA) of the central axis
of the streamers from which blobs are released, the total number
of blobs and the average daily rate released during the event, and
the minimum and maximum values of the blob velocities at a
specific height, say, 9 $R_\odot$. The PA increases
counterclockwise, taken to be zero in the northward direction. The
varying ranges of the deduced blob accelerations are also
presented in the last column of this table. The velocities and
accelerations given in this table are projected quantities on the
plane of the sky, which are obtained with a second-order
polynomial fitting to the measured blob tracks. The details of our
data reduction method will be introduced as we proceed.

\begin{table}[!htbp]
\caption{Information on the 10 events with quasi-periodic releases
of blobs lasting for three-four days observed in 2007.
 }

\begin{tabular}{clrccc}
  \hline
No.& Observation & PA ($^{\circ}$) & Total number & Velocity range
& Acceleration
\\ & date        &       &/avg. daily rate  &
at 9 $R_\odot$ (km s$^{-1}$) & range (m s$^{-2}$)
\\
  \hline
1 & Feb 14-16    & 103 & 11/3.7& 183-356 & 3.6-14.2 \\
2 & Apr 04-07    & 248 & 12/3  & 191-335 & 1.1-13.8 \\
3 & Apr 25-27    & 288 & 9/3   & 197-299 & 1.3-6.2  \\
4 & Apr 30-May 2 & 246 & 9/3   & 169-298 & 0.6-8.0  \\
5 & May 07-09    & 71  & 10/3.3& 240-400 & 3.6-18.2 \\
6 & Jun 05-08    & 67  & 13/3.3& 173-303 & 2.6-11.4 \\
7 & Jun 13-15    & 119 & 12/4  & 228-379 & 2.2-17.8 \\
8 & Jun 30-Jul 2 & 69  & 10/3.3& 162-287 & 2.2-11.2 \\
9 & Jul 18-20    & 106 & 11/3.7& 200-334 & 3.9-10.4 \\
10& Sep 27-29    & 244 & 9/3   & 192-286 & 2.2-11.0 \\
  \hline
\end{tabular}
\end{table}

The blob structures are only marginally brighter than the
background coronal emission, as seen from the white-light
brightness and polarization measurements by LASCO (Sheeley
\textit{et al}., 1997; Wang \textit{et al}., 1998). Therefore, it
is generally difficult to recognize a blob from the original
coronagraph images. The usual way to emphasize the blob features
is to make running-difference images by subtraction of two
successive images taken tens of minutes to one hour apart in time.
After this procedure, the blob structures are more easily
identified. They reveal themselves as radially elongated
white-leading-black bipolar islands. The white (black) color
indicates a brightness increase (decrease) in the corresponding
region during the elapsed interval. In the following discussion,
we first introduce our data analysis method by presenting two
examples observed during 13 to 15 June and during 30 June to 2
July, which are the seventh and eighth events listed in Table 1
and are referred to as Event A and Event B, respectively.

The two white-light images shown in Figures 1(a) and 1(b) are
recorded at 05:18 UT on 15 June and at the same time on 1 July,
where the white circle represents the surface of the Sun and the
one-quarter solid disk is where the LASCO C3 occulting disk is
located. The size of each image is 30 $R_\odot$ along the
horizontal direction and 15 $R_\odot$ along the vertical
direction. The standard routines provided with the solarsoft
software (http://www.lmsal.com/solarsoft/) are used to produce
these images. A background representing the contribution of the F
corona and instrumental stray light has been subtracted from each
image. It can be seen that a well-defined streamer exists at the
the southeastern part (PA$=$119$^{\circ}$) and the northeastern
part (PA$=$69$^{\circ}$) in Figures 1(a) and 1(b), respectively.
The blob structures that we are interested in are emitted right
atop of these two streamers. To recognize the blobs clearly, in
Figures 1(c) and 1(d) we present two running-difference images by
subtracting the images taken one hour earlier from those shown in
Figures 1(a) and 1(b). The blob structures are indicated with
white arrows. In Figure 1(d), two blobs are emitted successively
from the streamer. To view more blob events simultaneously in one
figure, we produce the temporal evolutionary map, which is the
stacked time series of radial strips centered along the
corresponding streamer stalk in the running-difference images.
This method has been used in previous blob studies (\textit{e.g.},
Wang \textit{et al}., 1998; Wang \textit{et al}., 2000). The width
of the narrow region is taken to be about 6 pixels, and the height
is given by the C3 FOV. Such height-time maps are presented in
Figures 1(e) and 1(f) for the two blob events, where the abscissa
represents the time of observation and the ordinate the height of
the strips.

It is obvious that the outward-moving blob structures are
represented as white-black tracks in these height-time maps. By
counting the number and deducing the slope of these tracks, we can
easily obtain the daily rate and the velocity profiles of the blob
structures. Note that only the data obtained by LASCO C3 are
analyzed in this paper. The reason for excluding the C2 data is
twofold. First, the blobs are observed initially near the streamer
tips, which are generally located at about 2 to 4 $R_\odot$ in the
middle part of the C2 FOV. At this height, it is generally
difficult to discern the blob structures even from the
running-difference images since the intensity of the background
streamer emission is relatively strong. The seeing condition of
blobs gets better when they enter the C3 FOV starting from 3.7
$R_\odot$. Second, the C3 FOV already covers the outer part of the
C2 FOV and our main purpose of this study is to search for the
persistent and quasi-periodic release events of blobs and to
determine the associated solar wind velocity, which is thought to
be well fulfilled by only using the C3 data.

As can be seen from Figures 1(e) and 1(f) there are a total of 12
blobs observed during the three days from 13 to 15 June and 10
blobs from 30 June to 2 July with an average daily rate being 4
and 3.3, respectively. By fitting the apparent blob tracks with a
second-order polynomial of the form
$r=r_{0}+v_{0}t+\frac{1}2{}at^{2}$, where $r_{0}$ and $v_{0}$
represent the heliocentric distance and speed at the starting
point of the selected event, the constant acceleration $a$ can be
determined by the quadratic fit. The temporal derivative of this
equation gives the expression of the fitted blob speed as
$v=v_{0}+at$.

The fitted velocity profiles as a function of heliocentric
distance are plotted in Figures 1(g) and 1(h) for the two events
discussed. Different symbols represent the velocities of different
blobs; the numbers before the symbols are ordered according to the
temporal sequence of the blob occurrence. As mentioned previously,
the blob speed can be used as a proxy for that of the mean solar
wind projected to the sky plane beyond a heliocentric distance of
about 5-7 $R_\odot$. We see that for most distances involved in
Figures 1(g) and 1(h) the symbols can be regarded as velocities
for both the blobs and the associated solar wind along the
streamer stalks. The velocities increase gradually with increasing
distances from 3.7 to 20 $R_\odot$. Also, it can be seen that the
velocities at a fixed distance vary significantly from blob to
blob. To indicate this, in Table 1 we present the varying ranges
of the blob velocities at 9 $R_\odot$ for all events. We see that,
for Event A, the minimum and maximum of the blob (or the solar
wind) velocities at 9 $R_\odot$ are 228 and 379 km s$^{-1}$,
respectively. The relative velocity variation is 66\% for this
event and 77\% for Event B.

To reveal more details of the velocity variability, in Figure 2 we
plot the fitted velocities at three heliocentric distances of 6
$R_\odot$ (squares), 9 $R_\odot$ (circles), and 12 $R_\odot$
(triangles) for Event A [Figure 2(a)] and B [Figure 2(b)]. The
abscissa of this figure is the time starting from 0 UT of the
first day of the event. It can be clearly seen that the speeds of
different blobs at a fixed distance vary significantly with time.
There are two possible physical causes accounting for such large
temporal velocity variations at a fixed distance. The first one is
the variation of the velocity of the local solar wind plasma, and
the second one is the change of the projection angle caused by
solar rotation during an event. We see that there is no apparent
regular pattern governing the velocity variations at the three
distances. Moreover, large velocity variations can take place
within a few hours. For example, for the first two blobs shown in
Figure 2(a) the velocity decreases abruptly from 430 to 280 km
s$^{-1}$ at 12 $R_\odot$ and from 355 to 242 km s$^{-1}$ at 9
$R_\odot$. The two blobs are separated temporally by several
hours. In such a short time, the effect of solar rotation on the
projection angle is basically negligible. Besides, if the temporal
change at a certain distance was caused by the projection effect,
the velocity would tend to be either monotonic or first increase
then decrease. Therefore, we suggest that the velocity change
presented in Figure 2 is mainly attributed to the velocity
variability of the local solar wind plasma. It is well known that
large velocity variability is one of the most apparent
characteristics of the slow solar wind (\textit{e.g.}, McComas
\textit{et al}., 2000). It has also already been mentioned in the
previous section that the wind along the plasma sheet above a
streamer is usually regarded as one source of the slow solar wind;
therefore, it is reasonable to deduce that this study provides an
observational manifestation of the large velocity variability of
the slow solar wind near the Sun.

Using exactly the same method of data reduction as that for these
two events, we examine the other eight events listed in Table 1.
The PA of the streamer axis, the total number and the average
daily rate of blobs in each event are shown in the third and
fourth columns of Table 1. The obtained height-time maps showing
the blob tracks are shown in Figure 3. The time of observation is
taken to be the abscissa, and the height of the radial strips
cropped from the series of running-difference images is shown as
the ordinate of this figure, the same as in Figures 1(e) and 1(f).
We can see that the most apparent common feature of the eight
panels in Figure 3 is the persistent and quasi-periodic
distribution of the white-black blob tracks. During each day of
these events, we observe about three to five blobs released along
the stalk of the corresponding streamer. The time span between two
adjacent blobs is about five to eight hours. Note that a coronal
mass ejection (CME) event was observed by LASCO C3 from 04:42 UT
on 9 May, whose white-black track is obviously brighter than that
of nearby blobs. For every blob release event, we have
double-checked the white-light images of LASCO C3 to determine
whether the white-black tracks in both Figure 3 and Figure 1 are
caused by the small-scale blob events or the large-scale eruptive
events. It is found that all the tracks are caused by small-scale
blob events except the one mentioned here, which is not included
in our statistics for the blobs.

In Figure 4, we plotted the velocity profiles of all the blobs
shown in Figure 3. The velocities are obtained by the same method
as that for Events A and B. Similarly to Figures 1(g) and 1(h),
the velocities of different blobs are represented with different
symbols, and the numbers in front of the symbols represent the
temporal order of the blob occurrence. It can be seen that, in all
these eight events, the blob velocities can vary significantly on
a time scale of several hours to a few days. Again, there are no
apparent patterns governing the velocity variations. For instance,
from 14 to 16 February with 11 blobs detected, the velocities at a
fixed distance, say, 9 $R_\odot$, vary dramatically in a few hours
from blob to blob. Specifically, the velocities of the first six
blobs are 356, 229, 313, 326, 253, and 223 km s$^{-1}$. As
suggested previously, such large velocity variability should be
taken as a consequence of the temporal evolution of the velocity
of the local slow solar wind. In other words, the data analysis
results shown in both Figure 1 and Figure 3 may provide
observational evidence for the presence of large velocity
variability of the slow wind near the Sun. Note that the varying
ranges of the fitted blob velocities at 9 $R_\odot$ have been
given in the fifth column of Table 1. In addition, the sixth
column of this table presents the minimum and maximum values of
the fitted acceleration, which also varies significantly from blob
to blob. From the analysis, we suggest that the slow solar wind
near the Sun flows outward from its source region already with
both a highly variable speed and acceleration. It is apparent that
both these two aspects may contribute to the large velocity
variability of the slow wind observed \textit{in situ} at much
greater distances. We point out in passing that the PAs of the 10
streamers used in this study are distributed over a wide range
from 67 to 288 degrees.

For all the blobs observed in the 10 events listed in Table 1, we
plot the velocity versus height profile in Figure 5. It can be
seen that the blobs generally accelerate gradually within the
LASCO C3 FOV. Their velocities increase slowly from 50-150 km
s$^{-1}$ at 3.7 $R_\odot$ and to 350-450 km s$^{-1}$ at 20
$R_\odot$. These statistical results are in ful agreement with
previous results by Wang \textit{et al}. (1998) using the data
observed near the last solar minimum. We expect more persistent
and quasi-periodic blob release events can be revealed in the
future.

\section{Summary and Discussion}

In this paper we have examined the LASCO C3 data obtained in 2007
and found 10 persistent and quasi-periodic blob release events
lasting for three to four days. The average daily rate of blobs is
found to be three to five, in agreement with previous studies for
the last solar minimum. It is found that the velocities of blobs
vary significantly from blob to blob over a time scale of several
hours to a few days. Taking the fitted blob speed beyond a certain
distance as a proxy for that of the mean flow, we suggest that the
large velocity variability, one of the most apparent signatures of
the slow solar wind observed \textit{in situ}, may develop near
the Sun, say, within the first tens of solar radii.

Sheeley, Wang, and coauthors (Sheeley \textit{et al}., 1997; Wang
\textit{et al}., 1998; Wang \textit{et al}., 2000) reported a few
persistent blob release events around the last solar minimum. To
interpret such steady blob releases from the tip of a streamer,
Chen \textit{et al}. (2009) proposed that the closed magnetic
field geometry associated with a streamer cusp can become unstable
to the expansion of the hot coronal plasmas, which results in a
so-called intrinsic instability of corona streamers and the
formation of blobs. For more details of this process, refer to the
first section of this paper or to Chen \textit{et al}. (2009). The
modeled number density and velocity signatures, even the daily
rate of blobs, are in agreement with previous observations.
However, it is also apparent that not all streamers are associated
with blobs. There are several possible reasons for this:
\textit{i}) The excitation and nonlinear development of the
mentioned instability require certain specific physical conditions
that develop over time. Blobs are not released, or, in other
words, the instability does not develop or develop maturely, if
the required conditions are not fulfilled or the developing
process is disturbed by other coronal activities such as CMEs.
\textit{ii}) The brightness of the blob structures is only
marginally higher than that of the background plasmas, so some
blobs, even if released, are not observable owing to the
limitations in resolution of current coronagraphs and interference
from instrumental backgrounds (\textit{e.g.}, stray light).
\textit{iii}) The blob signature may be obscured by other
structures or eruptive phenomena in the foreground or the
background corona along the line of sight.

The measurements of the dynamical parameters of the blob
structures provide important complements to the other
state-of-the-art techniques aimed at velocity diagnostics of the
solar wind near the Sun. It may be expected that a distribution
map of the solar wind velocities in the outer corona can be
coarsely delineated with enough data accumulated. Although the
flow velocity along the streamer stalk is provided only within a
height ranging from a few solar radii to about 20 $R_\odot$, it is
still useful for constraining the solar wind condition in the
outer corona. These constraints may help in establishing the
background conditions that can be used in models for CME
initiation and propagation, as well as for some space weather
forecasting models.

From Figure 2, we see that velocities of successive blobs at a
fixed distance can vary significantly within a few hours. Assuming
streamer blobs to be velocity tracers of the slow solar wind along
the plasma sheet, we therefore deduce that large velocity
variability, observed \textit{in situ} in the slow solar wind, is
already manifested near the Sun. It is a good question to ask how
the blob velocity variability compares with that of the slow solar
wind. To address this question, we examined the solar wind
velocity data obtained by, say, the \textit{Ulysses}/SWOOPS
instrument and found that large velocity variations similar to
that presented in Figure 2 are not unusual. However, we point out
that such comparisons should be conducted very carefully to reach
a physically meaningful conclusion. This is mainly due to the
large distance for the solar wind plasmas to travel from their
source region to the point of \textit{in situ} measurements. The
original velocity profiles as revealed by the blob observations
may undergo significant changes caused by the intrinsic dynamical
evolution and coupling processes with nearby solar wind plasmas.
The plasmas and magnetic structures associated with eruptive
transient events, such as magnetic clouds, may also contribute to
reshaping the solar wind velocity profiles. Therefore, comparison
between the blob variability and the slow wind variability is in
general not a trivial task, and so will not be further discussed
here.

Another very interesting and meaningful study would be to search
for the counterpart of the blob structures in interplanetary space
with \textit{in situ} data. As mentioned in the introduction,
there are models that suggest the blobs originate from
closed-field regions below the streamer cusp or along the current
sheet in the open magnetic geometry; therefore, the determination
of the \textit{in situ} blob counterpart will be helpful to
describe different formation mechanisms of blobs and to assess
plasma properties in the region near the streamer cusp. Many
spacecraft, such as \textit{Ulysses}, SOHO, \textit{Wind}, ACE, as
well as the recently launched STEREO (Kaiser \textit{et al}.,
2008; Galvin \textit{et al}., 2008), have already accumulated
enough data that would be appropriate for this study. The
\textit{in situ} counterpart of a blob could be recognized by
examining the elemental composition and abundance, ionic
temperature, and charge-state distribution, as well as the
magnetic-field geometry of the structures carried by the solar
wind. This study should be conducted in future.

\begin{acks}
The SOHO/LASCO data used here are produced by a consortium of the
Naval Research Laboratory (USA), Max-Planck-Institut f\"{u}r
Aeronomie (Germany), Laboratoire d'Astronomie (France), and the
University of Birmingham (UK). SOHO is a project of international
cooperation between ESA and NASA. This work was supported by
grants NNSFC 40774094, 40825014, 40890162, and NSBRSF
G2006CB806304 and by the Specialized Research Fund for State Key
Laboratory of Space Weather in China. H.Q. Song is grateful to
C.L. Shen, X.H. Zhao, and H.D. Chen for their assistance in
preparing this paper.
\end{acks}



\begin{figure}[tphb]
\includegraphics[width=1\textwidth]{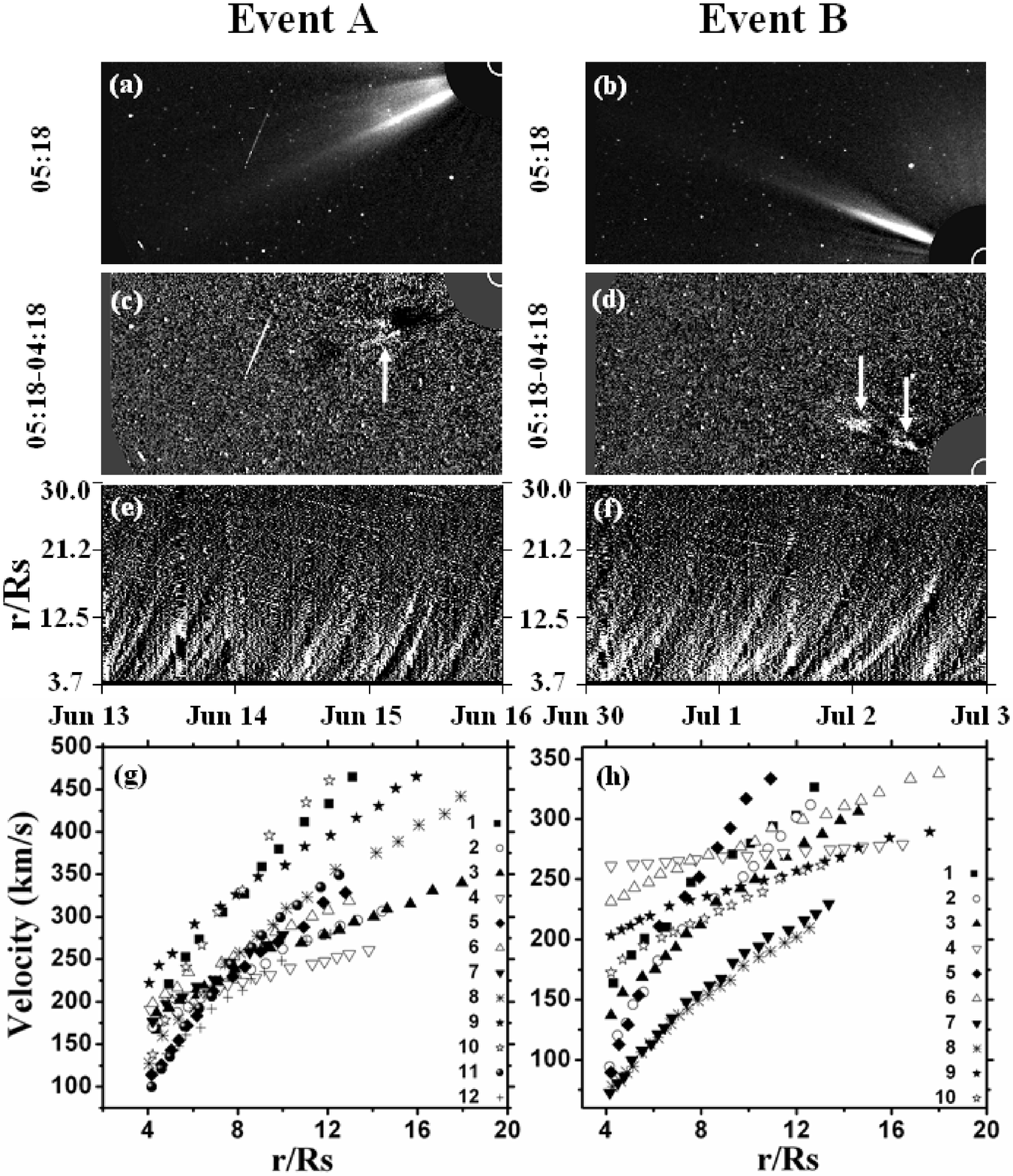}
\caption{Two examples of quasi-periodic releases of blobs observed
by LASCO C3, which are referred to as Event A (13-15 June ) and
Event B (30 June-2 July) in the text. In the eight panels, we
show: instantaneous, background-subtracted images recorded at 0518
UT on 15 June (a) and 1 July (b) by C3, cropped to 30 $R_\odot$ in
the horizontal direction and 15 $R_\odot$ in the vertical
direction; the difference of images taken at 0518 and 0418 UT on
15 June (c) and 1 July (d) with the same size as that of panels a
and b; height-time tracks of blobs for Events A (e) and B (f),
which are produced by stacking radial strips centered along the
streamer axis extracted from successive running-difference images;
and the fitted blob velocities as a function of heliocentric
distance (panels g and h for Events A and B). See text for more
details.}
\end{figure}

\begin{figure}[tphb]
\includegraphics[width=1\textwidth]{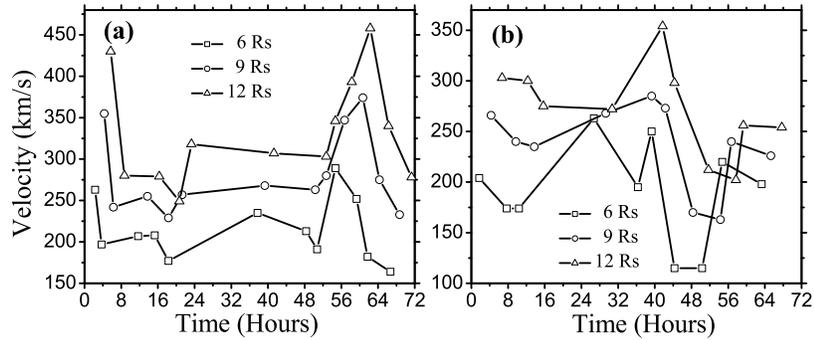} \caption{
The fitted velocities at three heliocentric distances 6 $R_\odot$
(squares), 9 $R_\odot$ (circles) and 12 $R_\odot$ (triangles) for
Events A (a) and B (b). The abscissa is the time starting from 0
UT of the first day of the event.}
\end{figure}

\begin{figure}[tphb]
\includegraphics[width=1\textwidth]{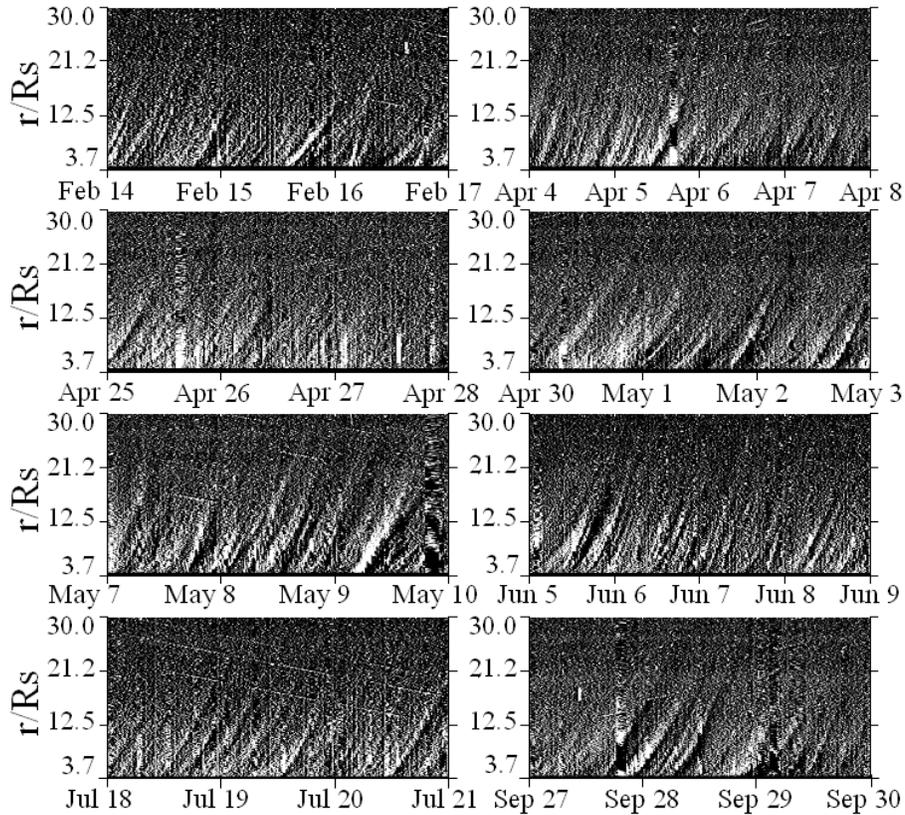}
\caption{Height-time tracks of blobs for the eight events listed
in Table 1. The images are produced by stacking radial strips
centered along the streamer axis extracted from successive
running-difference images of C3.}
\end{figure}

\begin{figure}[tphb]
\includegraphics[width=1\textwidth]{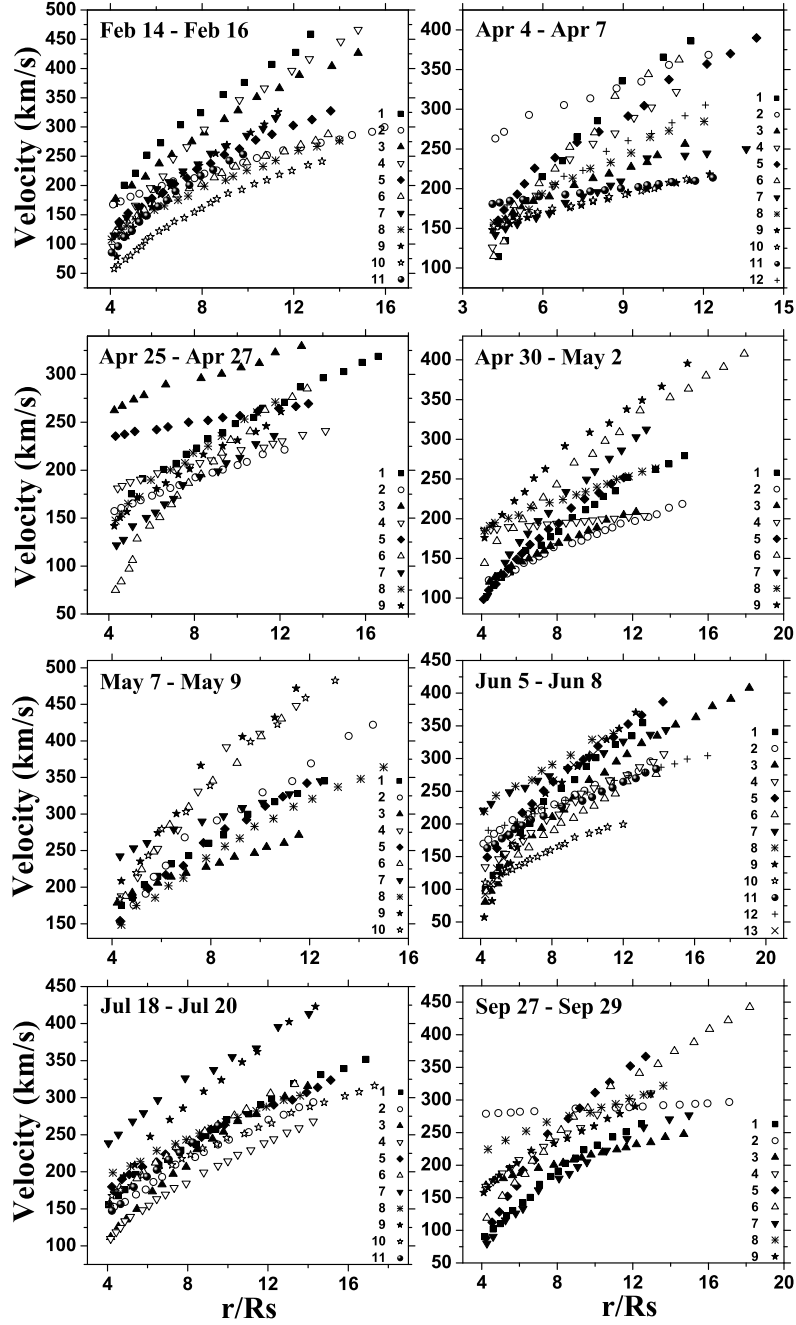}
\caption{The fitted velocities of blobs as a function of
heliocentric distance for the eight events listed in Table 1.}
\end{figure}

\begin{figure}[tphb]
\includegraphics[width=1\textwidth]{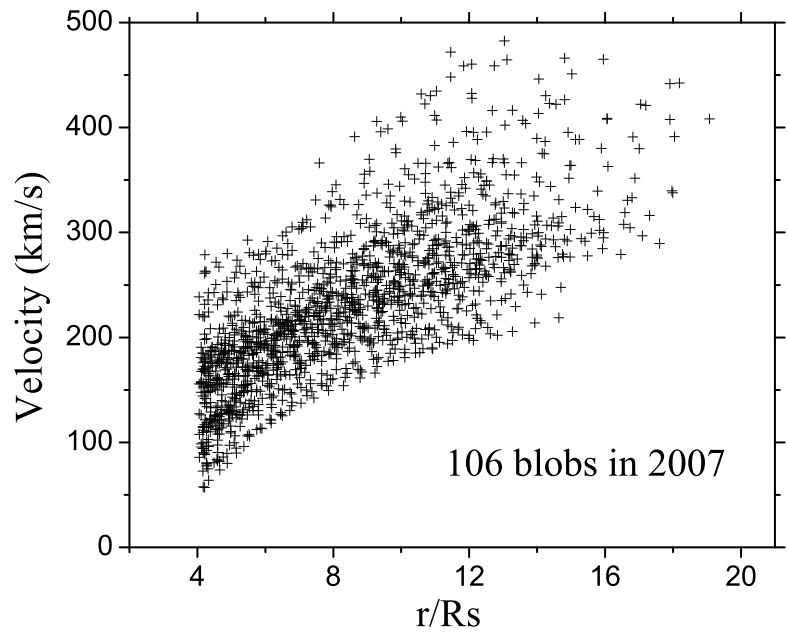}
\caption{Scatterplot of velocity versus heliocentric distance for
the 106 blobs observed in the 10 events listed in Table 1.}
\end{figure}

\end{article}

\end{document}